# Electrically Driven Spin Resonance of 4*f* Electrons in a Single Atom on a Surface

Stefano Reale[1,2,3], Jiyoon Hwang[1,4], Jeongmin Oh[1,4], Harald Brune[5], Andreas J. Heinrich[1,4], Fabio Donati[1,4]*, and Yujeong Bae[1,4]*

[1] Center for Quantum Nanoscience (QNS), Institute for Basic Science (IBS), Seoul 03760, Republic of Korea
[2] Ewha Womans University, Seoul 03760, Republic of Korea
[3] Department of Energy, Politecnico di Milano, Milano 20133, Italy
[4] Department of Physics, Ewha Womans University, Seoul 03760, Republic of Korea
[5] Institute of Physics, Ecole Polytechnique Fédérale de Lausanne, 1015 Lausanne, Switzerland

*Corresponding authors: F.D. (donati.fabio@qns.science), Y.B. (bae.yujeong@qns.science)

**A pivotal challenge in quantum technologies lies in reconciling long coherence times with efficient manipulation of the quantum states of a system. Lanthanide atoms, with their well-localized 4*f* electrons, emerge as a promising solution to this dilemma if provided with a rational design for manipulation and detection. Here we construct tailored spin structures to perform electron spin resonance on a single lanthanide atom using a scanning tunneling microscope. A magnetically coupled structure made of an erbium and a titanium atom enables us to both drive the erbium's 4*f* electron spins and indirectly probe them through the titanium's 3*d* electrons. In this coupled configuration, the erbium spin states exhibit a five-fold increase in the spin relaxation time and a two-fold increase in the driving efficiency compared to the 3*d* electron counterparts. Our work provides a new approach to accessing highly protected spin states, enabling their coherent control in an all-electric fashion.**

The last two decades have witnessed a rising focus on the control and application of quantum coherent effects, marking the advent of the so-called "second quantum revolution". Utilizing quantum coherent functionalities of materials for novel technologies, such as imaging, information processing, and communications, requires robustness of their quantum coherence, addressability, and scalability[1]. However, these requirements often clash since decoupling the quantum states from the environment prolongs the quantum coherent properties but hinders the possibility of efficient state manipulation.

Lanthanide atoms represent a promising platform to tackle this dilemma. Their well-localized 4f electrons show long spin relaxation $T_1$[2,3] and coherence times $T_2$[4,5]. In addition, their strong hyperfine interaction facilitates the read-out of nuclear spins[6,7]. In bulk insulators, exceedingly long $T_1$ and $T_2$ have been demonstrated using optical control and detection[8-11] down to the single atom level[12,13]. While hybrid optical-electrical approaches have been developed to access individual lanthanide atom's spins embedded in a silicon transistor[14], it is still challenging to achieve efficient control of the quantum states using electrical transport methods. This necessitates the rational design of a quantum platform capable of tackling both control and detection schemes, along with their interactions with local environments. In this context, single crystal surfaces constitute an advantageous framework both for building atomically engineered nanostructures and addressing individual spin centers, in particular using probe techniques[15-18]. However, coherent manipulation and detection of surface-adsorbed lanthanide atoms have so far remained elusive.

In this work, we demonstrate the control and detection of 4f electron spins by building atomic-scale structures on a surface using a scanning tunneling microscope (STM) with electron spin resonance (ESR) capabilities[19-22]. The atomic structures are composed of an erbium (Er) atom as the target spin system and a magnetically coupled titanium (Ti) atom as the sensor spin. This architecture allows us to drive ESR transitions on the Er 4f electrons with a projected angular momentum of ½[23] and to probe them indirectly through Ti. We observed an Er $T_1$ of close to 1 µs, 5 times longer than what was previously measured in 3d electrons with spin ½ on the same surface[18]. This novel platform allows for the ESR driving and read-out of the well-screened 4f electron spin states, paving the way to integrate lanthanide atoms in quantum architectures.

**Sensing Er Spin States through a Ti Atom**

Erbium atoms on a few monolayer-thick MgO(100) on Ag(100) present a $4f^{11}$ configuration with no unpaired electrons in the 5d and 6s shells[23]. The atomic-like spin and orbital momenta are coupled through the large spin-orbit interaction into a total angular momentum $\boldsymbol{J}_{\mathrm{Er}}$ with magnitude of $15\hbar/2$[23]. When adsorbed on the oxygen site of MgO (Fig. 1a), the crystal field leads to a strong hard-axis magneto-crystalline anisotropy that stabilizes a doubly-degenerate ground state with an out-of-plane component of the angular momentum $\pm\hbar/2$[23], which splits into two singlets when an external magnetic field (***B***) is applied. As found in a previous work[23], the component of $\boldsymbol{J}_{\mathrm{Er}}$ along the magnetic field direction (z), defined as $J_z$, increases from $\pm\hbar/2$ to $\pm 4\hbar$ by rotating ***B*** from the out-of-plane ($\vartheta = 0°$) to the in-plane ($\vartheta = 90°$) direction (Fig. 1b), while retaining a large probability for spin dipole transitions. Given these properties, Er can be regarded as a highly tunable two-level system allowing for efficient ESR driving. To characterize the magnetic states and anisotropy of Er, we utilized the dipole field sensing technique[24] with a Ti atom on the bridge binding site of MgO as a well-known spin sensor. On this binding site, Ti has a spin $S_{Ti}$ of magnitude ℏ/2 and a relatively weak g-factor anisotropy[25] with respect to the oxygen binding site[26].

We deposited Er and Ti at cryogenic temperatures (~10 K) on 2 monolayers of MgO grown on Ag(100) (Methods and Fig. S1a). Their binding sites on the surface can be changed by atom manipulation (Supplementary Section 2). Figure 1c shows the ESR spectra obtained on Ti in an Er-Ti dimer with the atomic separation of 0.928 nm (Fig. S2b). For $\vartheta$ = 8°, we observed one ESR peak at the resonance frequency of Ti which splits into two peaks separated by $\Delta f$ = 334 ± 3 MHz when rotating $B$ close to the in-plane direction ($\vartheta$ = 68°). The two ESR peaks stem from the magnetic interaction with the Er spin fluctuating between two states[24] during the measurement, with the relative peak intensity being proportional to the time-averaged population of the Er states. The pronounced difference in the relative intensity of the ESR peaks indicates a large imbalance in the Er state occupation even at $B$ = 0.3 T and 1.3 K, which reflects the large $J_z$ of Er at $\vartheta$ = 68° (Fig. 1b). The sign of this asymmetry depends on the character of the magnetic interactions between the two atoms. In Fig. 1c, the peak at the lower frequency is less intense than the one at the higher frequency and, hence, the interaction can be regarded as ferromagnetic[27].

The angle dependence of $\Delta f$ (Fig. 1d) gives a direct measurement of the Er-Ti interaction energy and of its anisotropy[24,27]. To interpret it, we model the system through a spin-Hamiltonian including both the single atom Zeeman and anisotropy terms, as well as the interaction between the two spins:

$$H = \mu_B g_{Er} \boldsymbol{B} \cdot \boldsymbol{J}_{Er} + D J_\perp^2 + \mu_B \boldsymbol{B} \cdot \mathbf{g}_{Ti} \cdot \boldsymbol{S}_{Ti} + H_{dip} + H_{exc}. \qquad (1)$$

Here, $\mu_B$ is the Bohr magneton, $J_\perp$ is the out-of-plane component of $\boldsymbol{J}_{Er}$, $g_{Er} = 1.2$ is the Er g-factor, and $\mathbf{g}_{Ti}$ is the Ti anisotropic g-tensor[25]. We use a magnetic anisotropy parameter $D = 2.4$ meV to match the Er energy splitting found in a previous study[23]. The magnetic coupling consists of dipolar ($H_{dip}$) and Heisenberg exchange interactions ($H_{exc}$):

$$H_{dip} = \frac{\mu_0 \mu_B^2}{4\pi \hbar^2 r^3}[g_{Er} \boldsymbol{J}_{Er} \cdot \mathbf{g}_{Ti} \cdot \boldsymbol{S}_{Ti} - 3(\hat{\boldsymbol{r}} \cdot g_{Er} \boldsymbol{J}_{Er})(\hat{\boldsymbol{r}} \cdot \mathbf{g}_{Ti} \cdot \boldsymbol{S}_{Ti})],$$

$$H_{exc} = \frac{J_{exc}}{\hbar^2}(\boldsymbol{J}_{Er} \cdot \boldsymbol{S}_{Ti}),$$

where $\mu_0$ is the vacuum permittivity, $r$ the separation between the two atoms, $\hat{r}$ the unit vector connecting them[24], and $J_{\text{exc}}$ the exchange interaction energy expressed in terms of $\boldsymbol{J}_{\text{Er}}$[28]. In our model, $J_{\text{exc}}$ is the only free parameter for the fit. As shown in Fig. 1d, our model accurately reproduces the data for $J_{\text{exc}}/\text{h} = -48$ MHz, where the negative sign indicates an antiferromagnetic coupling. This value is more than 20 times smaller than that observed for a Ti-Ti dimer at the same distance (–1.16 GHz)[29]. We ascribe the smaller Er-Ti coupling to the localization of the 4$f$ orbitals near the atom's core, which limits the overlap between Er and Ti orbitals when compared to the Ti-Ti case.

The strong angle dependence of $\Delta f$ can be understood by considering the large magneto-crystalline anisotropy of $\boldsymbol{J}_{\text{Er}}$. At $\vartheta = 90°$, $J_z$ is largest (4ℏ) and the angular momenta of both atoms are parallel to $\hat{r}$ (Fig. 1e), which maximizes the contribution of the dipolar coupling with a positive sign (ferromagnetic). When rotating $\boldsymbol{B}$ away from the in-plane direction, $\boldsymbol{S}_{\text{Ti}}$ follows the direction of $\boldsymbol{B}$, while the anisotropy of Er preserves a large component of $\boldsymbol{J}_{\text{Er}}$ mainly aligned along the in-plane direction (Fig. 1f). This misalignment between the two angular momenta reduces the dipolar interaction. Finally, as $\boldsymbol{B}$ approaches the surface normal (Fig. 1g), $\boldsymbol{J}_{\text{Er}}$ turns towards the out-of-plane direction with a much smaller value of $J_z = ℏ/2$. With the two momenta being perpendicular to $\hat{r}$, the dipolar contribution is minimal and negative (antiferromagnetic). Conversely, the mutual projection of $\boldsymbol{S}_{\text{Ti}}$ and $\boldsymbol{J}_{\text{Er}}$ is the only factor modulating the exchange interaction term, which remains negative (antiferromagnetic) at all angles.

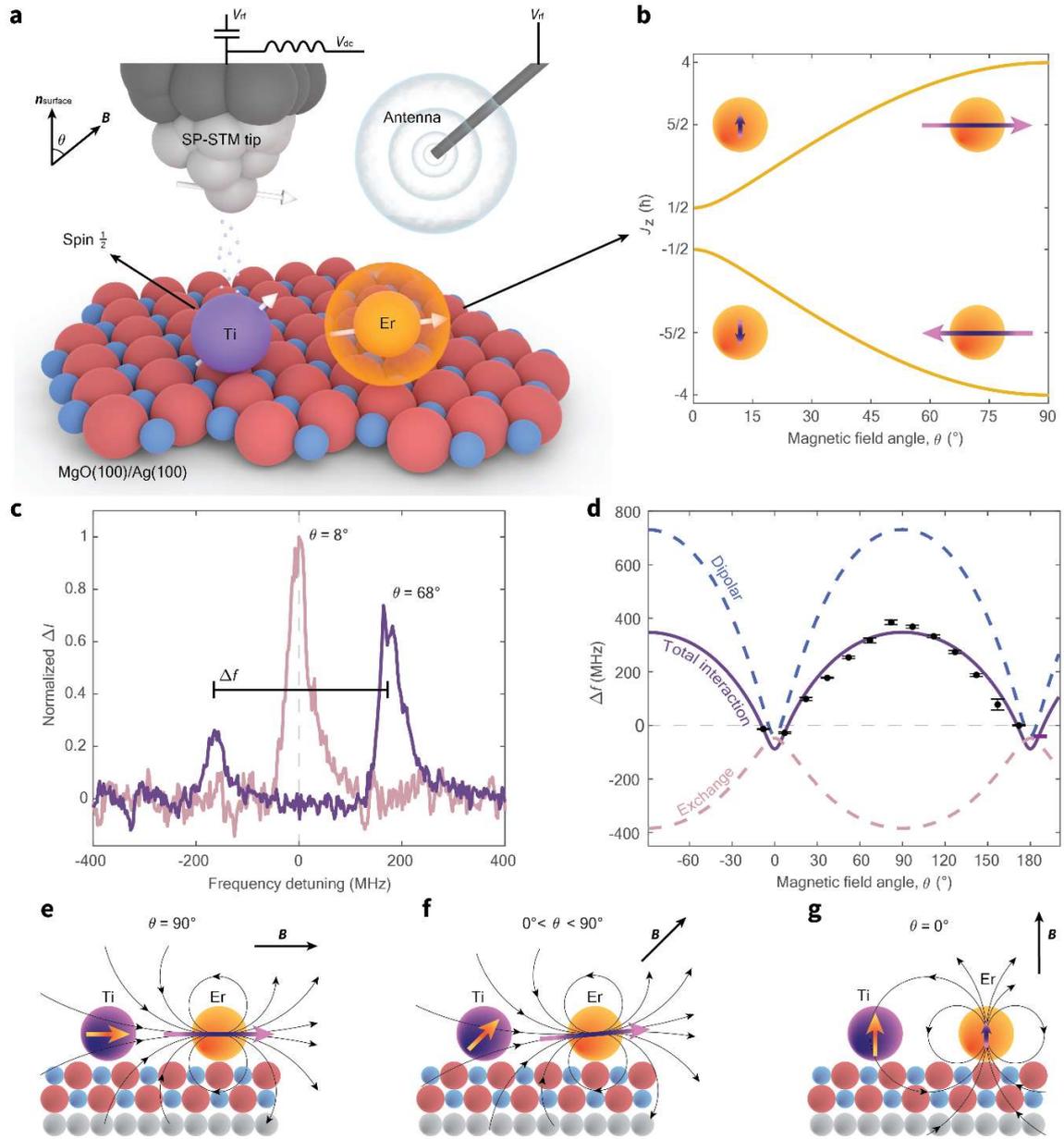

**Fig. 1 | Probing Er 4f electron spins through a Ti spin sensor. a**, Schematic of the experimental set-up for ESR-STM measurement of an Er-Ti dimer built on MgO/(100)/Ag(100). The Ti atom (purple) is positioned close to the Er atom (orange) and located under a spin-polarized (SP) STM tip. The external magnetic field (**B**) defines the z-direction and is applied at an angle $\vartheta$ from the out-of-plane direction. A dc voltage $V_{dc}$ is applied to the STM junction while the radio-frequency (rf) voltage is applied to the tip or to the antenna with an amplitude $V_{rf}$. **b**, The projected total angular momentum of Er ($J_z$) onto the **B** field direction as a function of $\vartheta$. The strong magnetic anisotropy favors an in-plane alignment of $J_{Er}$. **c**, ESR spectra of the Ti atom placed 0.928 nm apart from the Er atom at different $\vartheta$. At $\vartheta = 8°$, a single ESR peak is visible (pink) while, at $\vartheta = 68°$ (purple), the two ESR peaks are separated due to the magnetic interactions between the Er and Ti (set-point: $V_{dc}$ = 50 mV, $I_{dc}$ = 20 pA, $V_{rf}$ = 12 mV, B = 0.3 T). The spectrum at $\vartheta$ = 8° (pink) was normalized at its maximum intensity while the spectrum at $\vartheta$ = 68° (purple) was normalized to the sum of the intensities of its two peaks. The frequency detuning is defined with respect to 9.1 GHz (8.1 GHz) for the spectrum at $\vartheta = 8°$ ($\vartheta$ = 68°). **d**, ESR peak separation, $\Delta f$, as a function of $\vartheta$. The experimental points (black dots) were acquired at different set-points ($V_{dc}$

= 50 mV, $I_{dc}$ = 12-30 pA, $V_{rf}$ = 12-20 mV, $B$ = 0.3 T). The total interaction (solid purple line) calculated by the model Hamiltonian is composed of a dipolar contribution (dashed blue line) and an exchange contribution (dashed pink line). **e–g**, Schematic of the angular momenta of Er and Ti on MgO/Ag(100). The dipolar fields induced by Er are depicted as black curved arrows. When $B$ is applied along the in-plane direction ($\vartheta$ = 90°), the $J_z$ is maximum and aligned with the spin of Ti giving the largest ferromagnetic interaction. When $B$ is rotated, the spin of Ti follows the direction of $B$ while the total angular momentum of Er is aligned preferentially in-plane (**f**). In the out-of-plane direction ($\vartheta$ = 0°), $J_z$ is minimum and aligned with the spin of Ti (**g**) giving a small antiferromagnetic interaction.

**Spin Resonance of Er 4*f* Electrons**

The direct drive of ESR in STM requires positioning the tip directly on top of the target atom[19]. However, we observed no ESR when positioning the tip over an Er atom (Fig. S3b), which we attribute to the small polarization of the 5*d* and 6*s* shells of Er and to the weak interaction between the 4*f* and tunneling electrons. These factors were found to limit the tunneling magnetoresistance at the STM junction in other lanthanide atoms[30,31], possibly hindering both the ESR drive and detection[23].

To overcome this limitation, we built a strongly interacting Er-Ti dimer by positioning Ti at 0.72 nm from Er through atom manipulation (Fig. 2a and Supplementary Section 2). Similar to the isolated atom, we observed no ESR peaks at the Er position in the dimer (yellow curve in Fig. 2b). However, when the tip was positioned on Ti, we observed up to 5 peaks (pink curve in Fig. 2b). The first two peaks below 10 GHz with $\Delta f$ = 2.70 ± 0.01 GHz correspond to the ESR transitions of Ti that were similarly found in the dimer with larger atomic separations (Fig. 1c). Hence, we label them as $f_1^{\text{Ti}}$ and $f_2^{\text{Ti}}$, respectively. In this dimer, we observed that $f_1^{\text{Ti}}$ shows a higher intensity than $f_2^{\text{Ti}}$, indicating an antiferromagnetic exchange interaction[27] dominating over the dipolar coupling at this atomic separation. At higher frequencies, we further observed two peaks that are significantly blue-shifted when rotating $B$ from $\vartheta$ = 52° (pink curve in Fig. 2b) to 97° (purple). The higher resonance frequencies and pronounced angle dependence indicate that those transitions involve the large and anisotropic angular momentum of Er, and, thus, we label them as $f_3^{\text{Er}}$ and $f_4^{\text{Er}}$. In addition, their frequency separation exactly matches the one between $f_1^{\text{Ti}}$ and $f_2^{\text{Ti}}$, reflecting the same Er-Ti interaction. On the other hand, $f_3^{\text{Er}}$ and $f_4^{\text{Er}}$ are approximately equal in intensity, indicating that Ti fluctuates between two spin states with almost equal occupations. The comparable Ti states' occupation stems from the scattering with tunneling electrons and from the Zeeman splitting of Ti (~7 GHz) being smaller than the thermal energy at the measurement temperature of 1.3 K (~27 GHz). With $B$ at $\vartheta$ = 52°, we observed one more peak at even higher frequencies. Its frequency exactly matches the sum of $f_1^{\text{Ti}}$ and $f_4^{\text{Er}}$ (or equivalently $f_2^{\text{Ti}}$ and $f_3^{\text{Er}}$), which suggests an ESR transition involving both Ti and Er spins. We label this peak as $f_5^{\text{TiEr}}$. Remarkably, the sign of $f_3^{\text{Er}}$, $f_4^{\text{Er}}$ and $f_5^{\text{TiEr}}$ is opposite to that of $f_1^{\text{Ti}}$ and $f_2^{\text{Ti}}$, indicating a different detection mechanism for the transitions involving the Er spin, which will be discussed below. Finally, we observed an energy level crossing between Er and Ti transitions at $\vartheta \sim 12°$, with the Er resonance frequencies further shifting below the Ti transitions at $\vartheta \sim 0°$ (Fig. 2c and Fig. S4). This peculiar behavior is a consequence of the large difference in magnetic anisotropy between Er and Ti[23].

As shown in Fig. 2c, the angular dependence of the ESR frequencies is well reproduced by using Eq. 1 with $J_{\text{exc}}/h$ = –326 MHz. We observed small deviations for $f_1^{\text{Ti}}$, $f_2^{\text{Ti}}$ and $f_5^{\text{TiEr}}$, which we ascribe to different experimental conditions and magnetic interaction of Ti with the tip, which is not included in our model. Diagonalizing Eq. 1 allows us to analyze the quantum states of the Er-Ti dimer in terms of individual Er and Ti spin states. For an in-plane $B$ = 0.3 T, the energy detuning between the Er and Ti spins (30 GHz) is much larger than the interaction energy (about 3 GHz). Therefore, the Er-Ti dimer can be modeled with the 4 Zeeman product states of the Er and Ti spins. Following this picture, we can support the assignment of $f_{1,2}^{\text{Ti}}$ as Ti spin transitions occurring with no changes in the Er state, while $f_{3,4}^{\text{Er}}$ correspond to Er spin transitions without altering Ti. Finally, we attribute $f_5^{\text{TiEr}}$ to a double-flip transition involving both Er and Ti spins. Even though a $|\Delta m| > 1\hbar$ process is generally forbidden to first order, anisotropic terms in the magnetic interaction can give rise to higher order matrix elements connecting states with $\Delta m = \pm 2\hbar$[32].

When the field is oriented at $\vartheta = 0°$, both $\boldsymbol{J}_{\text{Er}}$ and $\boldsymbol{S}_{\text{Ti}}$ show an expectation value of $\hbar/2$, but a detuning still occurs due to the difference between the g-factors, $g_{\text{Er}}$ = 1.2 and $g_{\text{Ti}}$ = 1.989[25]. This detuning is comparable to their interaction energy and, thus, the two middle levels are no longer described by Zeeman product states (Fig. 2e). Finally, at the level crossing angle ($\vartheta \sim 12°$), the two Er and Ti middle levels become singlet and triplet states[29]. However, measuring ESR spectra under these conditions becomes challenging (Fig. S5), possibly due to the limitation in our detection as discussed in the following.

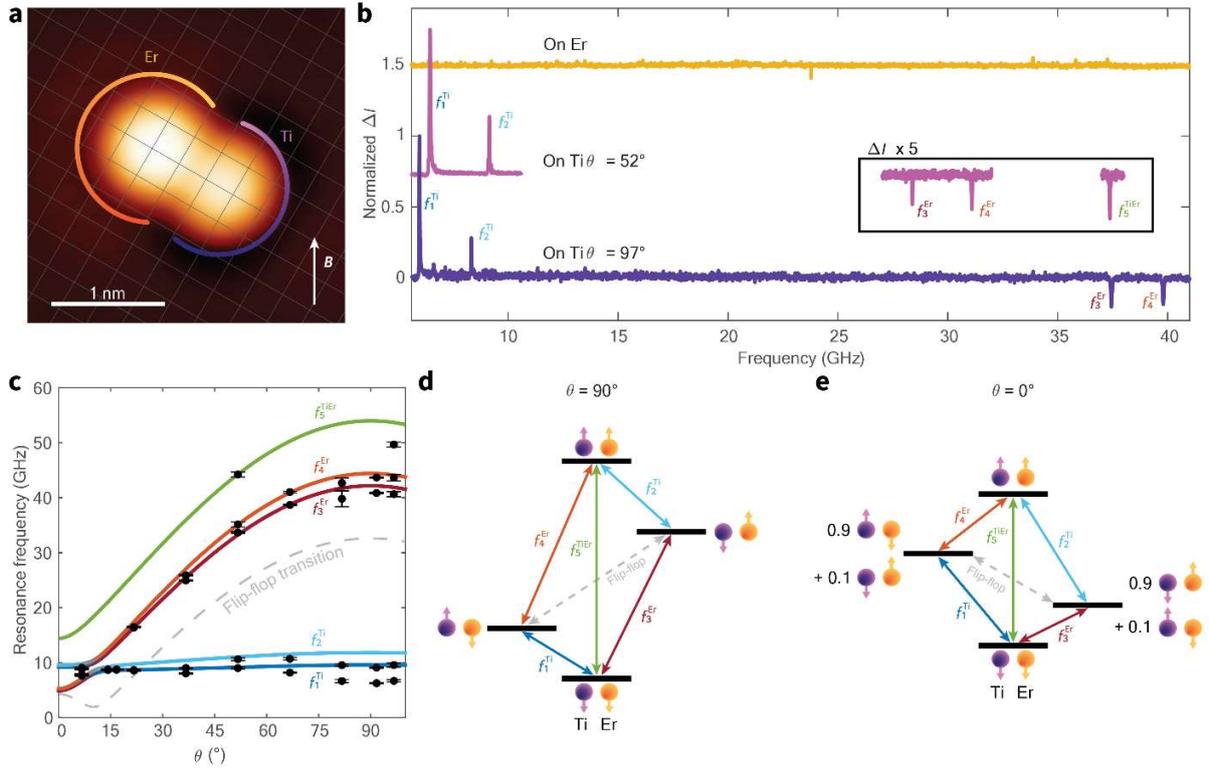

**Fig. 2 | Measurement of Er ESR transitions through a strongly coupled Ti atom. a**, Constant-current STM image of the engineered Er-Ti dimer with the atomic separation of 0.72 nm. The intersection of grids represents the oxygen sites of MgO. The Er atom (circled in yellow) is adsorbed on the oxygen site of MgO, while the Ti atom (circled in purple) is adsorbed on the bridge site (set-point: $V_{dc}$ = 100 mV, $I_{dc}$ = 20 pA). **b**, ESR spectra of the dimer given in **a**. When the STM tip is located on top of Er, no peaks are observed (yellow) (set-point: $V_{dc}$ = 50 mV, $I_{dc}$ = 20 pA, $V_{rf}$ = 20 mV, B = 0.28 T, $\vartheta$ = 97°). When the STM tip is located on top of Ti, 5 ESR peaks are detected ($f_{1,2}^{Ti}$, $f_{3,4}^{Er}$ and $f_5^{TiEr}$) with $\vartheta$ = 52° (pink), while 4 ESR peaks are detected ($f_{1,2}^{Ti}$, and $f_{3,4}^{Er}$) with $\vartheta$ = 97° (purple) (set-point: $V_{dc}$ = 70, 60 mV, $I_{dc}$ = 30, 40 pA, $V_{rf}$ = 20, 15 mV, B = 0.3 T). The spectra measured on Ti at $\vartheta$ = 52° and at $\vartheta$ = 97° were normalized at their respective maxima while the spectrum measured on top of Er was rescaled by the same amount used for the spectrum measured on Ti at $\vartheta$ = 97°. The spectra measured on Ti at $\vartheta$ = 52° and on Er are offset for clarity. **c**, ESR resonance frequencies as a function of $\vartheta$ at B = 0.32 T. The ESR frequencies obtained from each measurement are given as black dots alongside the transition energies predicted from the model Hamiltonian for $f_1^{Ti}$ (blue line), $f_2^{Ti}$ (light blue line), $f_3^{Er}$ (red line), $f_4^{Er}$ (orange line), $f_5^{TiEr}$ (green line) and flip-flop transition (dashed gray line). The experimental points were obtained at different set-points ($V_{dc}$ = 60–70 mV, $I_{dc}$ = 12–40 pA, $V_{rf}$ = 15–25 mV, B = 0.28–0.8 T); the resonance frequencies were rescaled by 0.32 T/B. **d,e**, Four-level schemes corresponding to the energies of the 4 spin states of the Er-Ti dimer and the corresponding transitions depicted as colored arrows at B = 0.32 T with different $\vartheta$ (90° and 0°, respectively). At $\vartheta$ = 90° (**d**) the spin states are given by the Zeeman products states, while at $\vartheta$ = 0° (**e**), a linear combination of the Zeeman product states is needed to describe the levels.

**Erbium ESR Detection and Driving Mechanisms**

The detection of ESR peaks exclusively occurs when the tip is positioned on top of Ti. Moving the tip from Ti to Er, the intensities of $f_3^{Er}$ and $f_4^{Er}$ gradually decrease and eventually vanish at ~0.3 nm from the Ti center (Fig. S6). This behavior indicates that driving an ESR transition on Er must induce a change in the Ti state occupation, subsequently modifying the spin polarization of the tunnel junction. In addition, Er ESR signals differ depending on specific tip conditions, i.e., different tips show positive or negative sign for $f_{3,4}^{Er}$ (Fig. 3a).

To further delve into the driving and detection mechanisms of the Er spin, we measured the intensities of $f_1^{Ti}$ and $f_3^{Er}$ as a function of $V_{rf}$ using a tip that shows negative Er peaks (Fig. 3b). While $f_1^{Ti}$ exhibits a continuous increase in intensity with increasing $V_{rf}$, $f_3^{Er}$ reaches saturation at $V_{rf}$ ~ 20 mV. The result for $f_1^{Ti}$ aligns with previous measurements on Ti[29], while the low-power saturation of Er is comparable to that of Fe, which might reflect a long $T_1$ and/or a high Rabi rate ($\Omega$)[33]. To understand this $V_{rf}$-dependence as well as the signs of ESR signals, we developed a rate equation model (Supplementary Section 7) based on the four-level scheme depicted in Fig. 3c. When driving $f_3^{Er}$, the populations of the initial and final states involved in the transition tend to equalize through a population transfer[34]. The changes in population are counteracted by the relaxation rates of each state ($\Gamma_{1,2}^{Ti}$ and $\Gamma_{3,4}^{Er}$), which tend to repopulate the depleted states. These rates are inversely proportional to the $T_1$ of the atom involved in the spin flip. Since Ti located under the tip is strongly influenced by tunneling electrons, relaxation events occur on a much shorter timescale than for Er[35], providing a more efficient pathway to attain the steady state. In addition, to account for the tip-dependent sign and intensity of Er ESR signals, we included a spin-pumping term originating from the spin-polarized current that can shift the Ti spin occupation (Fig. 3c for a negatively polarized tip)[17,36]. The proposed detection scheme based on the change of Ti state population accurately describes the $V_{rf}$-dependence (Fig. 3b) and the tip-dependent sign variations of the ESR signals (Fig. S7).

Finally, to identify the ESR driving source of the Er spin, we follow the relative peak intensity ($\Delta I/I_{dc}$) at different tip heights, as controlled by $I_{dc}$. As shown in Fig. 3d, $\Delta I/I_{dc}$ of $f_1^{Ti}$ increases with reducing the tip-sample distance, indicating that the main driving term for Ti arises from the exchange interaction with the spin-polarized tip[37,38]. On the other hand, $\Delta I/I_{dc}$ for $f_4^{Ti}$ remains independent of $I_{dc}$, which identifies the modulation of the magnetic interaction with Ti as the ESR driving source of Er[39]. The modulation of the magnetic coupling[40], in combination with anisotropic interaction terms[32], additionally explains the drive of the double-flip transition $f_5^{TiEr}$.

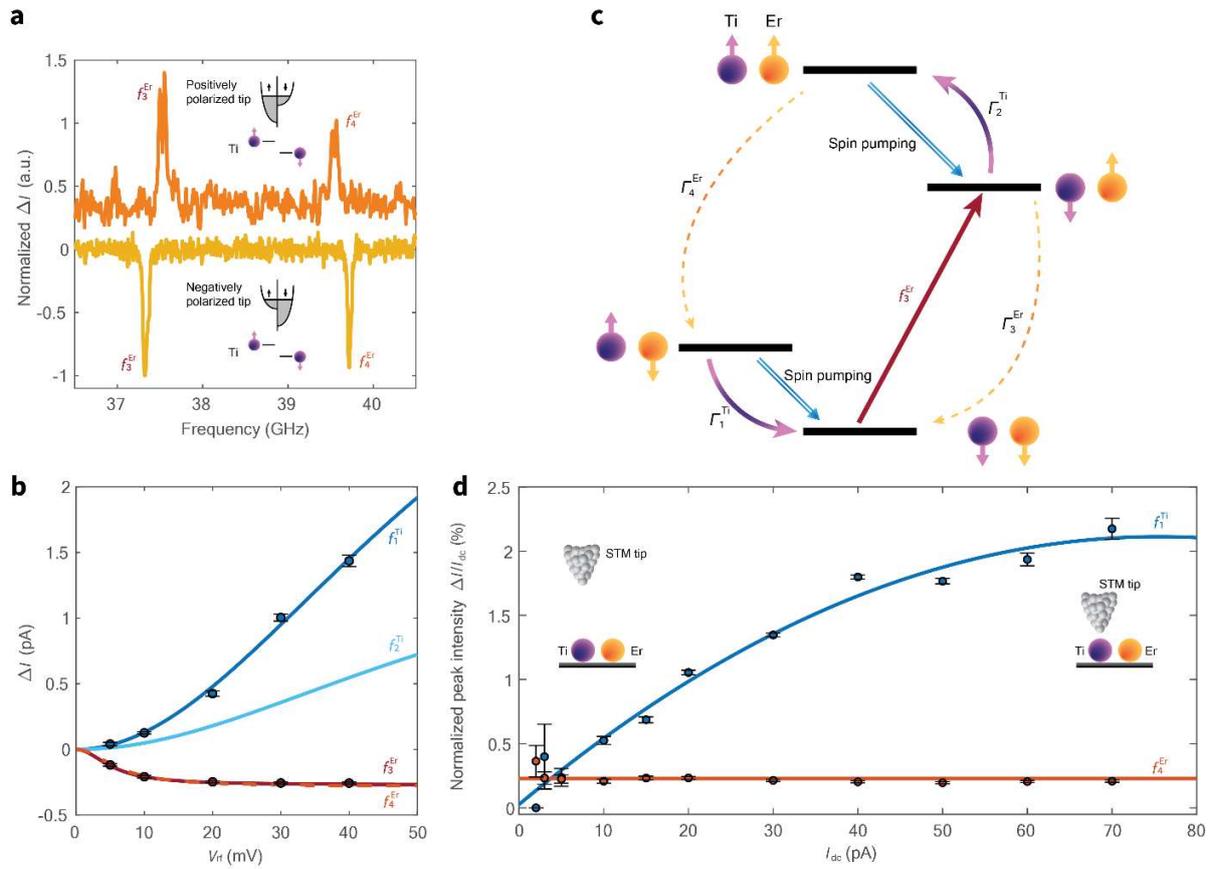

**Fig. 3 | Detection and driving mechanisms of Er ESR transitions. a**, ESR spectra showing $f_{3,4}^{Er}$ for two different STM tips: negative peaks related to negative spin-pumping (yellow line) and positive peaks related to positive spin-pumping (orange line) (set-point: $I_{dc}$ = 12, 20 pA, $V_{dc}$ = 70 mV, $V_{rf}$ = 25 mV, $B$ = 0.28, 0.32 T, $\vartheta$ = 67°). **b**, ESR peak intensities as a function of $V_{rf}$. The measured values for $f_1^{Ti}$ and $f_3^{Er}$ are given by black dots while the intensities predicted from the rate equation model for $f_{1,2}^{Ti}$ and $f_{3,4}^{Er}$ are given as blue, light blue, red solid lines and an orange dashed line, respectively (set-point: $I_{dc}$ = 40 pA, $V_{dc}$ = 70 mV, $B$ = 0.28 T, $\vartheta$ = 97°). **c**, Four-level scheme explaining the rate equation model while driving $f_3^{Er}$ (red arrow). The Ti's spin relaxation rates $\Gamma_1^{Ti}$ and $\Gamma_2^{Ti}$ are depicted as purple arrows while the Er's spin relaxation rates $\Gamma_3^{Er}$ and $\Gamma_4^{Er}$ are given as dashed yellow arrows. The negative spin pumping effect is represented as blue double arrows. **d**, Normalized ESR peak intensities ($\Delta I/I_{dc}$) for $f_1^{Ti}$ (blue circles) and for $f_4^{Er}$ (orange circles) at different tip heights. Here, the tip height is controlled by the set-point current $I_{dc}$, (set-point: $V_{dc}$ = 70 mV, $V_{rf}$ = 10 mV, $B$ = 0.28 T, $\vartheta$ = 97°). The blue and the orange lines serve as guides for the eye. The insets show two different tip-Ti distances: larger for low $I_{dc}$ and smaller for higher $I_{dc}$.

**Relaxation Time Measurement through Electron-Electron Double Resonance**

By applying an additional rf voltage ($V_{rf2}$), Ti and Er spins can be simultaneously driven in the so-called "electron-electron double resonance" scheme[41]. In a single-frequency ESR sweep, the relative intensities of $f_1^{Ti}$ and $f_2^{Ti}$ (Fig. 4a) reflect the thermal population of the Er spin. Instead, in double resonance the relative intensities of $f_1^{Ti}$ and $f_2^{Ti}$ are equalized when $f_3^{Er}$ is simultaneously driven (Fig. 4b). As shown in Fig. 4c, the intensity ratio of $f_1^{Ti}$ and $f_2^{Ti}$ ($\Delta I_{f_2}^{Ti}/\Delta I_{f_1}^{Ti}$) increases with increasing $V_{rf}$ only when $V_{rf2}$ is applied at the resonance frequency of $f_3^{Er}$ or $f_4^{Er}$, enabling selective modulation of the Er states to an out-of-equilibrium configuration.

Taking advantage of this selective driving mechanism, we implemented an inversion recovery measurement to estimate the spin relaxation time of Er ($T_1^{Er}$) in a pump-probe scheme (Fig. 4d). After exciting $f_3^{Er}$ with a pumping rf pulse of 200 ns duration that equalized the Er population, we applied a probe pulse of 500 ns for $f_1^{Ti}$ after a delay time $\Delta t$. Using this sequence, we monitored the time evolution of the intensity of $f_1$ as a function of $\Delta t$ from the out-of-equilibrium to the thermal state (Fig. 4e). The fit to an exponential function (Fig. 4e) gives $T_1^{Er}$ = 0.818 ± 0.115 μs, which is five times longer than what previously measured in Fe-Ti dimers in the absence of tunnel current[18]. We attribute this enhancement to the efficient decoupling of 4$f$ electrons from the environment, which reduces the relaxation events arising from the scattering with substrate electrons.

The large $T_1^{Er}$ measured through Ti indicates that the rapid spin fluctuations of Ti occurring on the timescale of a few ns[35] do not significantly perturb the stability of the Er states. This property partially originates from the large energy detuning between Er and Ti levels, which prevents the energy exchange required for spin-flip events. Using the experimentally obtained value of $T_1^{Er}$ in the rate equation model, we extract a driving term $W = \Omega^2 T_2/2$ for Er that is two times larger than for Ti in the same dimer (Supplementary Section 7). Despite the long spin lifetime and large driving term, attempts to drive Er Rabi oscillations through Ti do not yield a complete cycle (Fig. S8b), preventing a direct measure of the Er $T_2$. This is most likely due to a relatively low Rabi rate $\Omega$ provided by the moderate Er-Ti exchange coupling, which is about 2–3 times smaller than in the Fe-Ti dimer[39]. In turn, a low value of $\Omega$ together with a large driving term $W$ would imply much longer $T_2$ for Er than previous 3$d$ elements, highlighting the potential of 4$f$ electrons to realize higher performance atomic-scale qubits.

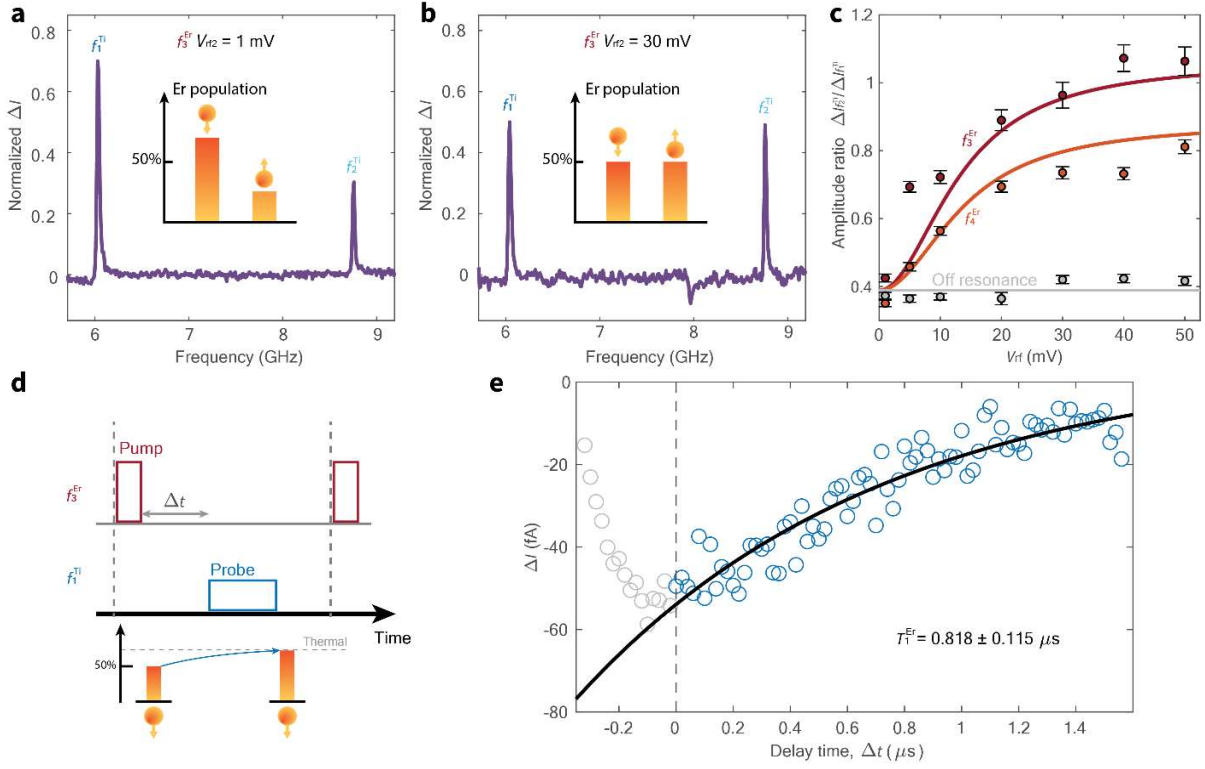

**Fig. 4 | Determination of Er spin relaxation time. a,b**, Double resonance spectra in the frequency range covering Ti ESR transitions $f_{1,2}^{Ti}$ (**a**) without and (**b**) with simultaneous driving of Er at the ESR frequency of $f_3^{Er}$. The peak intensities of $f_{1,2}^{Ti}$ are related to the relative population of the Er spin states (insets). The spectra were normalized to the sum of their peak intensity. **c**, ESR intensity ratios between $\Delta I_{f_2}^{Ti}$ and $\Delta I_{f_1}^{Ti}$ as a function of the driving strength $V_{rf2}$ at different Er ESR transition states (red, orange, and grey circles for $f_3^{Er}$, $f_4^{Er}$, and off-resonance, respectively). The solid curves show the correspondent simulation results by the rate equation model for $f_3^{Er}$ (red line), $f_4^{Er}$ (orange line) and at an off-resonance frequency (grey line). Set-point: $I_{dc}$ = 15 pA, $V_{dc}$ = 70 mV, $V_{rf}$ = 30 mV, $V_{rf2}$ = 1, 30 mV, $B$ = 0.28 T, $\vartheta$ = 97°. **d**, Schematics of the inversion recovery measurement in a pump-probe pulse scheme to determine the Er spin relaxation time $T_1^{Er}$. Each sequence is composed of a pump pulse at the resonance frequency of $f_3^{Er}$ (red box) and a probe pulse at the resonance frequency of $f_1^{Ti}$ (blue box). The probe pulse follows the pump pulse after a delay time Δ$t$. The population of the Er states after the pump pulse relaxes back to the thermal state following its $T_1$. **e**, The experimental data for the inversion recovery measurement (blue circles) show the intensity of the ESR signal at the probe pulse $f_1$ as a function of the delay time. The black line shows the fit using an exponential function with $T_1^{Er}$ of about 1 μs. Set-point: $I_{dc}$ = 50 pA, $V_{dc}$ = 70 mV, $V_{rf\,pump}$ = 60 mV, $V_{rf\,probe}$ = 100 mV, $B$ = 0.28 T, $\vartheta$ = 97°.

**Conclusions**

We demonstrated a new experimental approach to electrically drive ESR on the elusive *4f* electrons in a surface-adsorbed lanthanide atom with long spin relaxation time. Given the reduced scattering with the substrate electrons, it is reasonable to anticipate an enhancement in the coherence time of Er in comparison to 3*d* elements. We expect that, by employing a similar approach in different atomic structures, the ESR driving on the *4f* electrons can be amplified, enabling the use of lanthanide atoms as surface spin qubits with superior properties compared to the routinely adopted 3*d* elements.

**Methods**

**STM measurements**

Our experiment was performed in a home-built STM operating at the cryogenic temperature of ~1.3 K in an ultrahigh vacuum environment (< 1 x 10$^{-9}$ Torr)[42]. Using a two-axis vector magnet (6 T in-plane/4 T out-of-plane), the magnetic fields were varied from 0.28 T to 0.9 T at different angles from the surface[42]. To allow atom deposition on the sample kept in the STM stage, the sample is slightly tilted from the axis of the magnet by ~7° as estimated from the fit to the data shown in Fig. 1d. Considering this misalignment, all our experimental $\vartheta$ were offset by that amount accordingly. The magnetic tips used in our measurements were prepared by picking up ~4–9 Fe atoms from the MgO surface until the tips presented good ESR signals on isolated Ti atoms.

**ESR measurements**

We used two different schemes to apply $V_{rf}$ to the STM junction: one through the tip and one through an antenna (rf generators: Keysight E8257D and E8267D)[42]. In all our measurement involving a single rf sweep, we applied the $V_{rf}$ using an antenna located near the STM tip except for the data in Fig. 3b, where the $V_{rf}$ was combined with the dc bias voltage $V_{dc}$ using a diplexer at room temperature and then applied to the STM tip. The data in Fig. 4a–c were acquired by applying $V_{rf1}$ to the tip and simultaneously $V_{rf2}$ to the antenna. For the measurements reported in Fig. 4e and Fig. S8, the two rf voltages ($V_{rf1}$ and $V_{rf2}$) were combined through a power splitter (minicircuits ZC2PD-K0244+) and applied to the STM tip. For these measurements, both rf generators were gated by an arbitrary waveform generator (Tektronix, AWG 70002B).

**Sample preparation**

The surface of a Ag(100) substrate was cleaned by repeated cycles of Ar+ sputtering and annealing (700 K). We grew atomically thin layers of MgO(100) on the Ag(100) following a procedure described in a previous work[43]. We deposited Fe, Ti and Er atoms (< 1% of monolayer) from high purity rods (>99%) using an e-beam evaporator. During the deposition the sample was held at ~10 K in order to have well-isolated single atoms on the surface.

**Analysis of ESR spectra**

We fit the ESR spectra using a model given in[29] in order to extract the resonance frequency, peak intensity, and peak width for the data shown in Fig. 1d, Fig. 2c, Fig. 3b,d and Fig. 4c.

**Acknowledgements**

We thank Taehong Ahn and Leonard Edens for their support at the initial stage of the experiment and Yi Chen, Arzhang Ardavan, and Joaquín Fernández-Rossier for fruitful discussions. We acknowledge support from the Institute for Basic Science (IBS-R027-D1). Y.B. acknowledges support from Asian Office of Aerospace Research and Development (FA2386-20-1-4052). H.B. acknowledges funding from the SNSF AdG (TMAG-2_209266).

# Electrically Driven Spin Resonance of 4*f* Electrons in a Single Atom on a Surface

Stefano Reale[1,2,3], Jiyoon Hwang[1,4], Jeongmin Oh[1,4], Harald Brune[5], Andreas J. Heinrich[1,4], Fabio Donati[1,4]*, and Yujeong Bae[1,4]*

[1] Center for Quantum Nanoscience (QNS), Institute for Basic Science (IBS), Seoul 03760, Republic of Korea
[2] Ewha Womans University, Seoul 03760, Republic of Korea
[3] Department of Energy, Politecnico di Milano, Milano 20133, Italy
[4] Department of Physics, Ewha Womans University, Seoul 03760, Republic of Korea
[5] Institute of Physics, Ecole Polytechnique Fédérale de Lausanne, 1015 Lausanne, Switzerland
*corresponding authors: F.D. (donati.fabio@qns.science), Y.B. (bae.yujeong@qns.science)

**Table of contents**

### 1- Experimental set-up and identification of atomic species

From the STM image in Fig. S1a, it is possible to distinguish the MgO(100) patch from the Ag(100) substrate by the different apparent height. On top of the MgO patch different atoms are distinguishable by their distinct apparent heights: ~130 pm for Ti on the oxygen site ($Ti_O$), ~210 pm for Ti on the bridge site ($Ti_B$), ~210 pm for Er on the oxygen site ($Er_O$), ~285 pm for Er on the bridge site ($Er_B$), and ~170 pm for Fe on the oxygen site. The species are further identified by their d$I$/d$V$ spectra (Fig. S1b–k). While most of the atoms are distinguishable from the apparent heights and the spectral features, the $Er_O$ and $Ti_B$ present a similar apparent height as well as no clear spectral features. To distinguish these two species, we utilize the spin-polarized STM tip. In contrast with $Er_O$ (Fig. S1h), the d$I$/d$V$ spectrum on $Ti_B$ measured using the spin-polarized STM tip (Fig. S1g) presents a feature characteristic of a spin-flip excitation at around 0 mV (*1*) similarly to $Ti_O$ (Fig. S1f). In the main text and in the following sections, we simply refer to $Ti_B$ as Ti and $Er_O$ as Er.

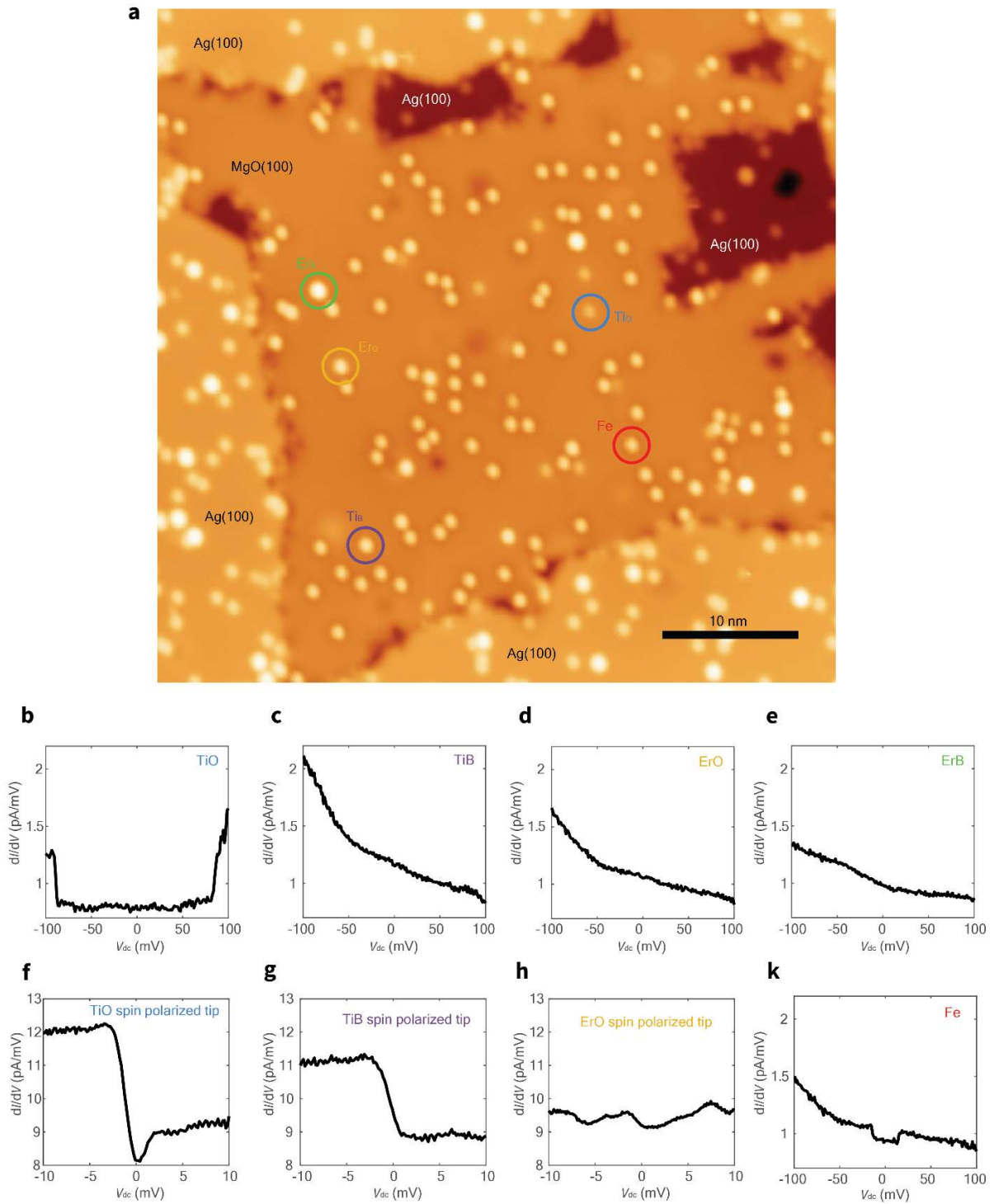

**Figure S1 | Characterization of the atomic species. a**, Constant current STM image of the Ag(100) surface partially covered by two-monolayers of MgO(100) (set point: $V_{dc}$ = 100 mV, $I_{dc}$ = 20 pA). The different atomic species can be distinguished by their apparent heights and d$I$/d$V$ features: (**b**) Ti on the oxygen site (Ti$_O$), (**c**) Ti on the bridge site (Ti$_B$), (**d**) Er on the oxygen site (Er$_O$), (**e**) Er on the bridge site (Er$_B$), Fe (**k**). (**f**) Ti$_O$ and (**g**) Ti$_B$ present a spin-flip excitation at around 0 mV when measured with a spin polarized tip, while no excitation is present on Er$_O$ (**h**).

## 2- Atom manipulation to construct Er-Ti dimers

The Er-Ti dimer used to acquire the data presented in Fig. 1c,d was built through atom manipulation. After identifying a Ti atom and an Er atom, we manipulated the Ti adsorption site by the following procedure: 1) position the STM tip 1 lattice site away from the Ti center (set point: $V_{dc}$ = 100 mV, $I_{dc}$ = 20 pA), 2) switch off the STM feedback, 3) approach the tip by 330 pm to the surface, 4) apply a voltage pulse of 330 mV, and 5) switch the feedback on. This procedure allows us to move the Ti atom by half lattice sites (from $Ti_O$ to $Ti_B$ and vice versa) in a controlled manner. We repeated this procedure until we obtained the desired Er-Ti dimer with a distance of 0.928 nm (Fig. S2) with the Ti atom placed at the (-2, 2.5) lattice position from Er. We used a similar procedure to preapare the Er-Ti dimers with 0.72 nm separation with the Ti atom placed at the (±2.5, 0) or (0, ±2.5) lattice position from Er.

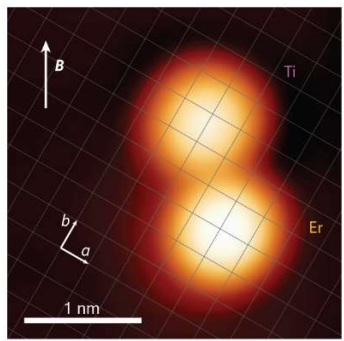

**Figure S2 | Constant-current STM image of the Er-Ti dimer with 0.928 nm separation** (set point: $V_{dc}$ = 100 mV, $I_{dc}$ = 20 pA). The intersection of grids represents the oxygen site of MgO and the lattice vectors (*a*,*b*) are superimposed on the grid.

## 3- Electron spin resonance on isolated Ti and Er

In order to perform ESR on an isolated Er atom, we confirmed whether the prepared spin-polarized tip is suitable to perform ESR or not, by measuring the ESR signal on an isolated Ti atom (Fig. S3a). When positioning the same tip over an isolated Er atom, however, no ESR peak was detectable (Fig. S3b)

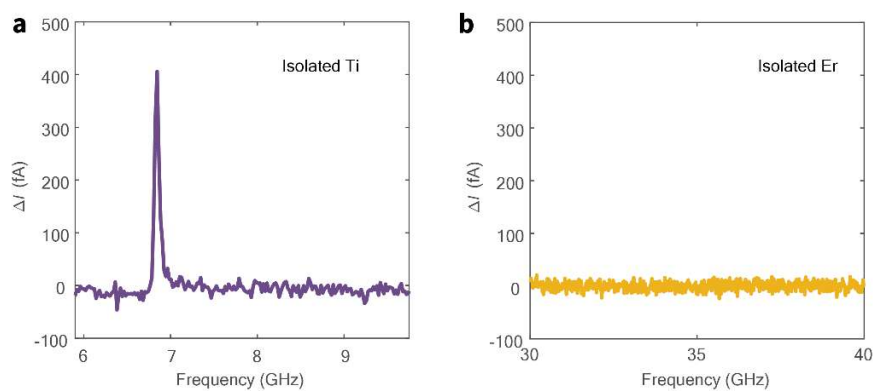

**Figure S3 | Electron spin resonances on the isolated atoms. a**, ESR spectrum measured with the STM tip positioned above an isolated Ti atom (set point: $V_{dc}$ = 60 mV, $I_{dc}$ = 20 pA, $V_{rf}$ = 20 mV, $B$ = 0.28 T, $\vartheta$ = 97°). **b**, ESR spectrum measured with the STM tip positioned above an isolated Er atom (set point: $V_{dc}$ = 50 mV, $I_{dc}$ = 50 pA, $V_{rf}$ = 20 mV, $B$ = 0.28 T, $\vartheta$ = 97°).

## 4- Electron spin resonance spectra with out-of-plane magnetic fields

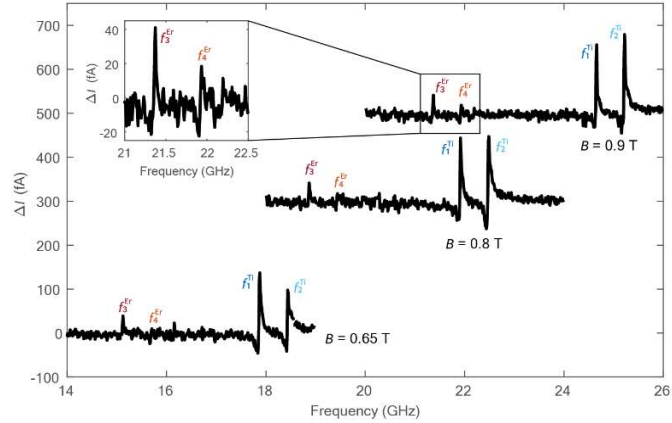

**Figure S4 | Electron spin resonance at out-of-plane magnetic fields.** ESR spectra measured on the Er-Ti dimer with 0.72 nm separation at different magnetic fields close the out-of-plane direction ($\vartheta = 7°$): $B$ = 0.65 T, 0.8 T, and 0.9 T (set point: $V_{dc}$ = 70 mV, $I_{dc}$ = 20 pA, $V_{rf}$ = 25 mV). The spectra at $B$ = 0.8 T and 0.9 T were shifted in $\Delta I$ by 300 fA and 500 fA, respectively, for clarity. The inset on the top left corner shows a zoomed-in spectrum of the peaks $f_3$ and $f_4$.

When we applied the magnetic field close to the out-of-plane direction ($\vartheta = 7°$), we observed 4 ESR peaks (Fig. S4). As explained in the main text, when the magnetic field is applied out-of-plane ($\vartheta = 0°$) the expectation value of $\boldsymbol{J}_{\mathrm{Er}}$ is $\hbar/2$ similarly to $\boldsymbol{S}_{\mathrm{Ti}}$. However, the Er g-factor is 1.2, while Ti has a g-factor of 1.989 (*3*). The peaks at lower frequencies, thus, correspond to the Er ESR transitions ($f_3^{\mathrm{Er}}$ and $f_4^{\mathrm{Er}}$) due to the smaller Zeeman energy of Er than the one of Ti. To further clarify the identification of ESR peaks, we measured the ESR speactra at different magnitudes of magnetic fields at $\vartheta = 7°$ and followed the linear dependence of their resonance frequencies on the magnetic field magnitudes.

## 5- Electron spin resonance spectra close to the level crossing

When the magnetic field is applied at about 12° from the normal to the surface ($\vartheta \sim 12°$), we expect Er and Ti to have similar Zeeman splittings. In this situation, the intermediate energy levels of the Er-Ti dimer with 0.72 nm separation must be regarded as singlet ($\frac{1}{\sqrt{2}}|\uparrow\Downarrow\rangle - \frac{1}{\sqrt{2}}|\uparrow\Downarrow\rangle$) and triplet states ($\frac{1}{\sqrt{2}}|\uparrow\Downarrow\rangle + \frac{1}{\sqrt{2}}|\uparrow\Downarrow\rangle$) (*4*) as explained in the main text. In Fig. S5, we show three ESR spectra acquired around the expected matching angle, i.e. $\vartheta$ = 14.5°, 17°, and 22°. When the ESR spectrum is acquired at $\vartheta$ = 22°, all four ESR transitions from $f_1^{\mathrm{Ti}}$ to $f_4^{\mathrm{Er}}$ are visible, with $f_3^{\mathrm{Er}}$ and $f_4^{\mathrm{Er}}$ observed at higher frequencies than $f_1^{\mathrm{Ti}}$ and $f_2^{\mathrm{Ti}}$. In addition, the different peak intensities between $f_1^{\mathrm{Ti}}$ and $f_2^{\mathrm{Ti}}$ (with the intensity of $f_1^{\mathrm{Ti}}$ larger than $f_2^{\mathrm{Ti}}$) indicate an antiferromagnetic coupling between Er and Ti (*5*). Conversely, for both $\vartheta$ = 17° and $\vartheta$ = 14.5°, it is not possible to identify the $f_3^{\mathrm{Er}}$ and $f_4^{\mathrm{Er}}$ peaks stemming from Er ESR transitions. Nevertheless, for both spectra measured for $\vartheta \leq 17°$ $f_1^{\mathrm{Ti}}$ and $f_2^{\mathrm{Ti}}$, the asymmetry is reversed, suggesting a change of the system configuration possibly due to the close match between the Er and Ti levels at around $\vartheta$ = 12°. As discussed in the main text, close to $\vartheta$ = 12° the energy levels of the system cannot be represented as Zeeman product states and for this reason the detection mechanism for both the Ti and the Er peaks explained in the main text may not be valid in this range of $\vartheta$.

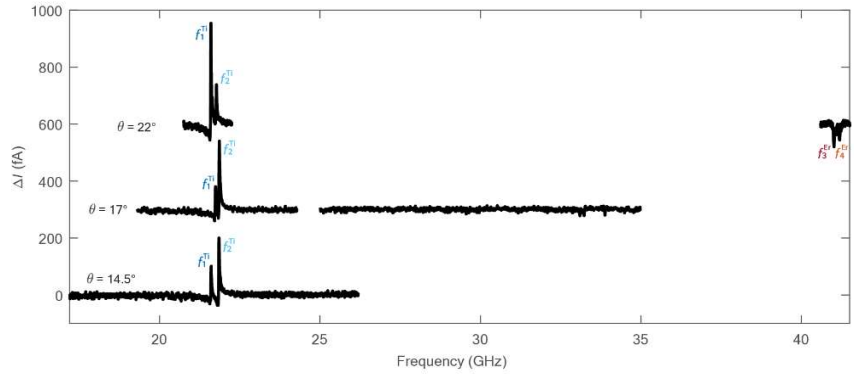

**Figure S5 | Electron spin resonance close to the level crossing point.** ESR spectra on the Ti atom of the 0.72 nm dimer at $\vartheta$ = 14.5°, $\vartheta$ = 17° and $\vartheta$ = 22° (set point: $V_{dc}$ = 70 mV, $I_{dc}$ = 12 pA, $V_{rf}$ = 20 mV, $B$ = 0.8 T). The spectra at $\vartheta$ = 17° and $\vartheta$ = 22° were shifted in $\Delta I$ by 300 fA and 600 fA respectively for clarity.

### 6- Tip position dependence of ESR signals on the Er-Ti dimer

As mentioned in the main text, when the tip is positioned above the Ti atom in the Er-Ti dimer with 0.72 nm separation, we can resolve up to 5 ESR peaks. However when we move the tip away from the Ti center, the peaks related to the Er ESR transitions decrese in intensity (Fig. S6). When the tip is about 0.3 nm from the Ti center the intensity of the peaks is too low to be resolved (spectrum 2). In a similar way, when the tip is positioned above Er no peaks are detectable (spectrum 1).

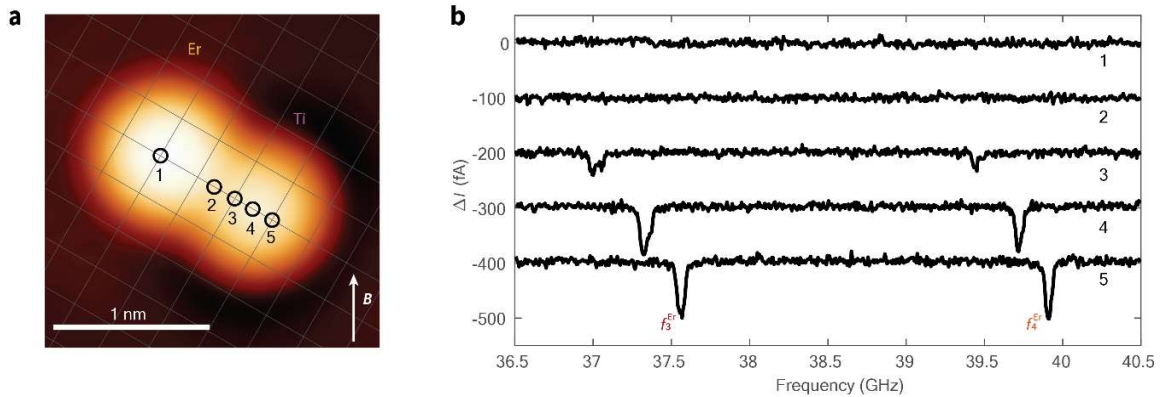

**Figure S6 | Electron spin resonance at different tip locations on the Er-Ti dimer. a**, STM image of the Er-Ti dimer with spacing of 0.72 nm (set point: $V_{dc}$ = 100 mV, $I_{dc}$ = 20 pA) with a grid superimposed representing the MgO lattice. The different locations where the tip was positioned during the ESR measurement are depicted as black circles numbered from 1 to 5. **b**, ESR spectra at different tip locations on the dimer (set point: $V_{dc}$ = 70 mV, $I_{dc}$ = 12 pA, $V_{rf}$ = 25 mV, $B$ = 0.28 T, $\vartheta$ = 97°), when the tip is approached laterally to the Er atom, the Er ESR peaks, $f_3^{Er}$ and $f_4^{Er}$, shift to lower frequencies due to the antiferromagnetic interaction between the Er atom and the magnetic tip. The intensity of the peaks decreases with moving the tip away from the Ti atom and no peaks are detectable at a distance of 0.3 nm from its center (spectrum 2). The spectra 2, 3, 4 and 5 were shifted vertically by -100 fA, -200 fA, -300 fA and -400 fA respectively for clarity.

### 7- Rate equation model

To reproduce the different sign of the peaks of the Er's ESR transitions shown in Fig. 3a, as well as the dependence of the peak intensity as a function of the driving strength ($V_{rf}$) displayed in Fig. 3b of the main text, we developed a rate equation model based on the four-level scheme picture of Fig. S7a. The rate equations consist of 4 differential equations:

$$\frac{dn_{00}}{dt} = -(W_1 + \Gamma_{1-}^{Ti})n_{00} - (W_3 + \Gamma_{3-}^{Er})n_{00} + (W_1 + \Gamma_{1+}^{Ti})n_{10} + (W_3 + \Gamma_{3+}^{Er})n_{01}$$

$$\frac{dn_{10}}{dt} = -(W_1 + \Gamma_{1+}^{Ti})n_{10} - (W_4 + \Gamma_{4-}^{Er})n_{10} + (W_1 + \Gamma_{1-}^{Ti})n_{00} + (W_4 + \Gamma_{4+}^{Er})n_{11}$$

$$\frac{dn_{01}}{dt} = -(W_2 + \Gamma_{2-}^{Ti})n_{01} - (W_3 + \Gamma_{3+}^{Er})n_{01} + (W_2 + \Gamma_{2+}^{Ti})n_{11} + (W_3 + \Gamma_{3-}^{Er})n_{00}$$

$$\frac{dn_{11}}{dt} = -(W_2 + \Gamma_{2+}^{Ti})n_{11} - (W_4 + \Gamma_{4+}^{Er})n_{11} + (W_2 + \Gamma_{2-}^{Ti})n_{01} + (W_4 + \Gamma_{4-}^{Er})n_{10}$$

Here, $n_x$ is the population of the level $x$, $W_y$ is the driving of the transition $y$, and $\Gamma_y$ is the relaxation rate of the levels connected by the transition $y$. The "+" and "–" superscripts in the relaxation rates indicate if the relaxation is towards a lower energy level (+) or a higher energy level (–), such that the total relaxation can be written as $\Gamma_y = \Gamma_{y+} - \Gamma_{y-}$. We distinguish the total relaxation rates for Er and Ti by computing $\Gamma^{Er} = 1/T_1^{Er}$ and $\Gamma^{Ti} = 1/T_1^{Ti}$. In addition, we conserve the total population, $n_{00} + n_{10} + n_{01} + n_{11} = 1$. To obtain the steady state solution we set the derivatives equal to zero. By considering a specific driving $W_y$ and solving the system of equations in the steady state, we obtain the population of each level. We included the spin pumping term ($\zeta$) as a shift of the population of the levels given by the injection of $|\uparrow\rangle$ states into Ti, as follows:

$$n_{00}^{\text{spin pumping}} = n_{00} - \zeta$$

$$n_{10}^{\text{spin pumping}} = n_{10} + \zeta$$

$$n_{01}^{\text{spin pumping}} = n_{01} - \zeta$$

$$n_{11}^{\text{spin pumping}} = n_{11} + \zeta$$

We used the rate equation model to fit the $V_{rf}$ dependence of the peak intensity in Fig. 3b. We set $T_1^{Ti}$ = 10 ns (7) and $T_1^{Er}$ = 818 ns. We used two different driving terms for Er and Ti transitions: $W^{Ti} = W_1 = W_2 = A^{Ti}V_{rf}^2$ and $W^{Er} = W_3 = W_4 = A^{Er}V_{rf}^2$, where $A^{Ti,Er}$ is the respective scaling factor for the driving of Ti and Er ESR. The fitting parameters are $A^{Ti}$, $A^{Er}$ and the spin pump term. The fitting yielded $A^{Ti}$ = 14917, $A^{Er}$ = 29338 and a negative spin pump of –0.94%. The driving term can be expressed as $W = T_2\Omega^2/2$ (8). The rabi rate $\Omega$ depends on the strength of the modulation provided by the tip-atom or atom-atom coupling and it is linear with $V_{rf}$ and, thus, the scaling factor $A$. As discussed in the main text and in the previous section, the Er-Ti coupling is 3–4 times smaller than Fe-Ti dimers used for remote ESR of 3d electrons (6, 9). Therefore, we expect a lower Rabi rate for Er since its driving comes from the modulation of its magnetic interaction with Ti. On the other hand, the fit yields a driving factor $A^{Er}$ larger than $A^{Ti}$, which suggests a much longer $T_2$ of Er compared to other 3d elements, possibly due to the well protected 4f orbitals.

The spin pumping term $\zeta$ is required to account for the different sign of the Er peaks observed with different ESR tips, as shown in Fig. 3a in the main text. As discussed in the following, our model indicates that a negative $\zeta$ produces negative Er peaks while the opposite is true for a positive $\zeta$. In Fig. S7b, we show the effect of $\zeta$ on $f_3^{Er}$ in the rate equation model. In the absence of spin pumping, a small positive signal is predicted. This is due to the difference in energy between the ESR transitions $f_1^{Ti}$ and $f_2^{Ti}$ which leads to an intrinsic difference between $\Gamma_1^{Ti}$ and $\Gamma_2^{Ti}$.

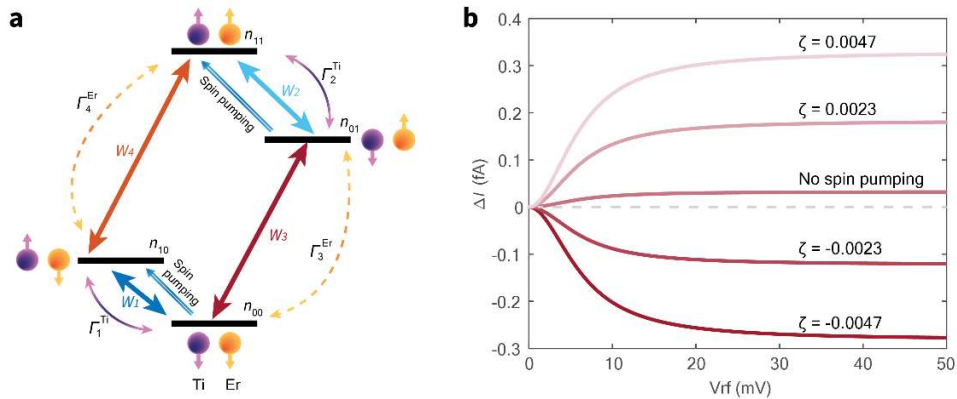

**Figure S7 | Rate equation model and spin pumping dependence of the signal. a**, 4-level scheme relative to the rate equation model reporting the driving term ($W$), relaxation rates ($\Gamma$) and a positive spin pump term ($\zeta$). **b**, Effect of the spin pump term for the $f_3^{Er}$ peak intensity ($\Delta I$) as a function of $V_{rf}$ predicted by the rate equation model. A negative $\zeta$ produces a negative ESR signal while a positive $\zeta$ produces a positive ESR signal. When the spin pumping is excluded from the model, a slightly positive ESR signal is predicted.

## 8- Measurement of Rabi oscillations

To measure Rabi oscillations on Ti and Er atoms, we followed a procedure similar to the one used in (*7*). With the STM tip positioned on top of the Ti atom in the Er-Ti dimer with 0.72 nm separation, we applied a series of $V_{rf}$ pulses at the resonance frequency of $f_1$ (Fig. S8a) and $f_3$ (Fig. S8b) with increasing pulse widths. We subtracted a linear fit to the data in order to remove the rf rectified current given by the nonlinearity of the *I-V* curve (*10*). When we apply rf pulses at the resonance frequency of $f_1$ we can resolve Rabi oscillations in the Ti spin (Fig. S8a). The fit of the signal measured on a Ti atom with an exponentially decaying sinusoidal function yields a Rabi rate $\Omega_{Ti}$ of 435 MHz ± 41 MHz and a $T_{2\,Rabi}^{Ti}$ of 9.9 ns ± 3.4 ns. On the other hand, when we apply rf pulses at the resonance frequency of $f_3$, no Rabi oscillation is observed. The monotonic decrease of the signal is due to the negative sign of the $f_3$ peak, which reaches saturation for sufficiently long rf pulses.

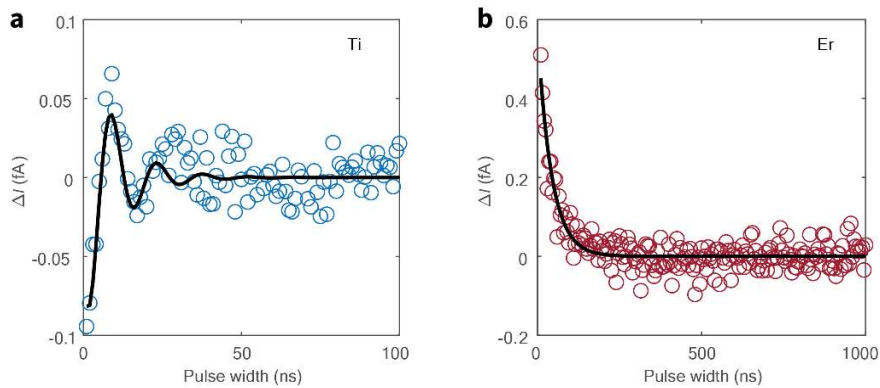

**Figure S8 | Rabi measurements. a**, Pulsed ESR measured with the STM tip on the Ti atom of the Er-Ti dimer with 0.72 nm separation. With the pulses applied at the resonance frequency of $f_2^{Ti}$, Rabi oscillations for the Ti spin are clearly observed (set point: $V_{dc}$ = 20 mV, $I_{dc}$ = 10 pA, $V_{rf}$ = 70 mV, B = 0.288 T, $\vartheta$ = 97°). The black line is a fit using an exponentially decaying sine function. **b**, Pulsed ESR measured with the STM tip on the Ti atom of the same dimer but with the pulses at the resonance frequency of $f_3^{Er}$ (set point: $V_{dc}$ = 70 mV, $I_{dc}$ = 50 pA, $V_{rf}$ = 90 mV, B = 0.288 T, $\vartheta$ = 97°). The black line is an exponential fit as a guide for the eye.

## 9- Dimer with $^{167}$Er

When measuring different Er-Ti dimers at the same separations (0.72 nm), we observed that a small fraction of them do not show any peak in the Er ESR transition range (Fig. S9). We ascribe this observation to the presence of $^{167}$Er isotopes on the surface, which is the only observationally stable isotope of Er with a non-zero nuclear spin, present with a 22.9% abundancy. This isotope presents a nuclear spin of 7ℏ/2. When driving ESR transitions on this atom, a single ESR peak is expected to split into 8 peaks due to its hyperfine interaction with the nuclear spin (*11*). However, as for this atom the intensity of the Er ESR transition is also reduced by a factor of 8 and, the ESR signal becomes too small to be detected in the present detection scheme. Nevertheless, this observation further supports the interpretation that $f_3$ and $f_4$ correspond to ESR transitions in the Er 4*f* spins since these transitions are the only ones that should be affected by the hyperfine interaction between the Er electron and nuclear spins.

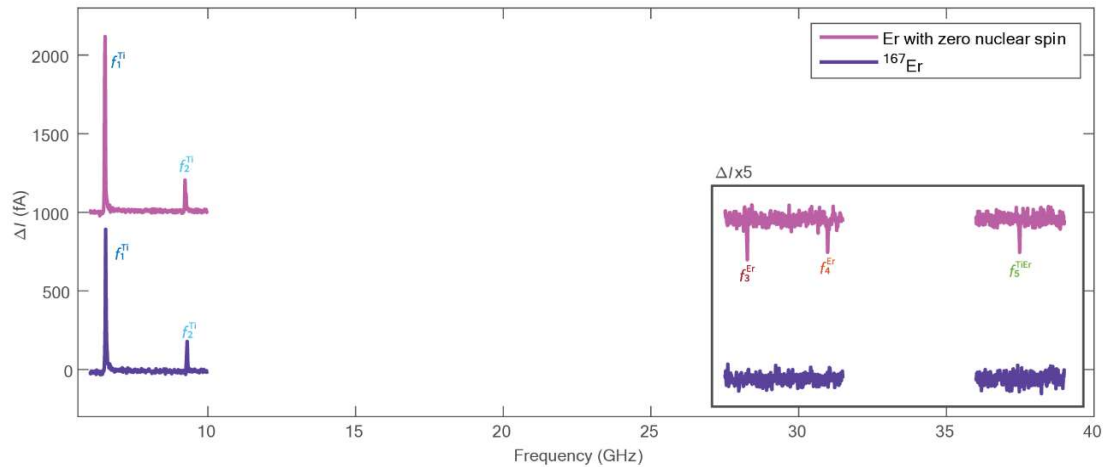

**Figure S9 | Electron spin resonance on dimer containing $^{167}$Er.** ESR spectra acquired with the same STM tip on top of Ti in two different dimers: in the standard Ti-Er dimer (pink line, shifted in Δ$I$ by -1000 fA for readability) 5 peaks are detectable ($f_1^{Ti}$, $f_2^{Ti}$, $f_3^{Er}$, $f_4^{Er}$ and $f_5^{TiEr}$) while in the dimer containing $^{167}$Er with nuclear spin 7ℏ/2 (purple line), only $f_1^{Ti}$ and $f_2^{Ti}$ (related to Ti transitions) are detectable (set point: $V_{dc}$ = 60 mV, $I_{dc}$ = 20 pA, $V_{rf}$ = 15 mV, $B$ = 0.3 T, $\vartheta$ = 52°).